\documentclass[10pt]{article}
\usepackage[a4paper,margin=2.5cm]{geometry}
\usepackage[utf8]{inputenc}
\usepackage[T1]{fontenc}
\usepackage{authblk}
\usepackage{graphicx}
\usepackage{natbib}
\usepackage[french, english]{babel}
\usepackage{csquotes}
\usepackage{lineno}
\usepackage{amsmath}
\usepackage{amsfonts}
\usepackage{textcomp}
\usepackage{gensymb}
\usepackage{amssymb}
\usepackage{bm}
\usepackage{setspace}
\usepackage{tabularx}
\usepackage{adjustbox}
\usepackage{multirow}

\ifdefined\correction
  \newcommand{\add}[1]{{\bf \color{blue}{#1}}}
  \newcommand{\rem}[1]{{\it \color{red}{#1}}}

  \newcommand{\remli}[1]{{\it \color{red}{#1}}}
\else
  \newcommand{\add}[1]{#1}
  \newcommand{\rem}[1]{}

  \newcommand{\remli}[1]{}
\fi 

\usepackage{hyperxmp}
\usepackage{hyperref}
\hypersetup{colorlinks,linkcolor={blue},citecolor={blue},urlcolor={red},breaklinks,draft=false,
pdfauthor={Marc Peruzzetto}, 
pdftitle={Simulation des écoulements gravitaires avec les modèles d'écoulement en couche mince : état de l'art et exemple d'application aux coulées de débris de la Rivière du Prêcheur (Martinique, Petites Antilles)}}  

\DeclareUnicodeCharacter{03BC}{$\mu$}

\title{Simulation des écoulements gravitaires avec les modèles d'écoulement en couche mince : état de l'art et exemple d'application aux coulées de débris de la Rivière du Prêcheur (Martinique, Petites Antilles)}

\author[1, 2]{Marc Peruzzetto}
\author[1]{Gilles Grandjean}
\author[2]{Anne Mangeney}
\author[1]{Clara Levy}
\author[1]{Yannick Thiery}
\author[3]{Benoit Vittecoq}
\author[4]{François Bouchut}
\author[5]{Fabrice R. Fontaine}
\author[2]{Jean-Christophe Komorowski}

\affil[1]{BRGM, F-45060 Orléans, France}
\affil[2]{Université Paris Cité, Institut de physique du globe de Paris, CNRS, F-75005 Paris, France}
\affil[3]{BRGM, F-97200 Fort-de-France, Martinique}
\affil[4]{Laboratoire d’Analyse et de Mathématiques Appliquées (UMR 8050), CNRS, Univ. Gustave Eiffel, UPEC, Marne-la-Vallée, F-77454, France}
\affil[5]{Observatoire volcanologique et sismologique de la Martinique, Institut de physique du globe de Paris, F-97250 Fonds Saint Denis, France}

\begin{document}

\selectlanguage{french}

\flushbottom
\maketitle
\thispagestyle{empty}

\section*{Résumé}
La quantification de la propagation des glissements de terrain est une étape clé de l'analyse des risques gravitaires. Dans ce contexte, les modèles d'écoulement\rem{s} en couche mince sont de plus en plus utilisés pour simuler la dynamique d'écoulements gravitaires comme les coulées de débris. Ils sont plus souples d'utilisation et moins coûteux en temps de calcul que des modèles 3D\add{ plus complexes}, et fournissent des informations plus précises sur les vitesses et les épaisseurs des écoulements que des méthodes purement empiriques. Dans cette revue de la littérature, nous présentons les principales rhéologies utilisées pour modéliser des écoulements gravitaires homogènes, et donnons un exemple d'application pratique avec la Rivière du Prêcheur (Martinique, Petites Antilles). Nous discutons ensuite les principales pistes de développements permettant d'utiliser ces modèles dans le cadre d'études opérationnelles d'analyse d'aléas et de risques.

\selectlanguage{english}

\section*{Abstract}

Quantifying the propagation of landslides is a key step for analyzing gravitational risks. In this context, thin-layer models have met a growing success over the past years to simulate the dynamics of gravitational flows, such as debris flows. They are easier to use and require less computing ressources than 3D models, and provide more detailed estimations of flow thicknesses and velocities than purely empirical methods. In this litterature review, we present the main rheologies used to model homogeneous gravitational flows, and describe a practical case study with debris flow modeling in the Prêcheur River (Martinique, Lesser Antilles). Then, we discuss possible developments for operational hazard and risk assessment with thin-layer models. 

\selectlanguage{french}

\section*{Introduction}

D'après \citet{petley_global_2012} et \citet{froude_global_2018}, 55~000 personnes ont été tuées dans le monde par des \rem{glissements}\add{mouvements} de terrain entre 2004 et 2016, \rem{hors}\add{sans compter les} \rem{glissements}\add{mouvements} déclenchés par des séismes. \rem{Ces glissements sont également très meurtriers mais la quantification du nombre de victimes est dans ce cas plus complexe \citep{petley_global_2012}.}Les \rem{glissements}\add{mouvements} de terrain peuvent également avoir un impact économique significatif\add{. Par exemple, en Espagne, les coûts directs (réparation et remplacement des infrastructures ou bâtiments affectés) sont estimés à 200 millions de dollars par an \citep{kjekstad_economic_2009}. En France, }\rem{comme }le glissement du Chambon en 2015 a provoqué la ferm\add{e}t\rem{e}ure de la route reliant Grenoble à Briançon pendant 2 ans \citep{dubois_glissement_2016}. Si les zones montagneuses sont les plus touchées par ces phénomènes, les régions c\add{ô}tières et les reliefs moins escarpés peuvent aussi être affectés. Ainsi, les \rem{côteaux}\add{coteaux} de la vallée de la Marne et de la Montagne de Reims connaissent régulièrement des déstabilisations qui endommagent les axes de communication ainsi que des parcelle\add{s} cultivées \citep[e.g.][]{deneeckhaut_comparison_2010}.

\rem{Compte tenu de ces enjeux économiques et humains, la quantification des risques gravitaires peut être un élément clé des politiques d'aménagement du territoire. Les études de risques nécessitent alors en premier lieu de cartographier les aléas en répondant à plusieurs questions qui présentent toutes des verrous scientifiques importants \citep{fell_guidelines_2008, corominas_recommendations_2014}:
\begin{enumerate}
    \item Quel type d'événement est susceptible de se produire sur le site d'étude?
    \item Quand les glissements peuvent-ils se produire?
    \item Où ces événements peuvent-ils être initiés? 
    \item Après la rupture, jusqu'où les glissements peuvent-ils se propager? 
    \item Quelle est l'intensité des phénomènes attendus (e.g. vitesse, épaisseur, volume)
\end{enumerate}}
\add{Compte tenu de ces enjeux économiques et humains, la communauté internationale \citep{fell_guidelines_2008, corominas_recommendations_2014} préconise une stratégie en 4 étapes pour analyser l'aléa puis évaluer les risques associés :
\begin{enumerate}
	\item Caractérisation du type de mouvements de terrain pouvant se produire sur le site d'étude. Cela passe souvent par la constitution d'un inventaire;
	\item Établissement d'une carte de susceptibilité par type de phénomène, intégrant la probabilité spatiale d'occurrence des événements (i.e., identification des zones susceptibles de générer des mouvements de terrain) et la probabilité d'atteinte compte tenu des zones de départ identifiées. Dans la littérature, la notion de susceptibilité se rapporte parfois seulement à la probabilité spatiale d'occurrence;
	\item Établissement d'une carte d'aléa (par type de phénomène ou pour tous les types de mouvements de terrains), qui intègre les probabilités spatiales et temporelles d'occurrence des phénomènes, selon leur magnitude (e.g. volume) et/ou leur intensité (e.g. vitesse). 
	\item Analyse des risques, c'est à dire de l'impact des mouvements de terrain sur des enjeux (population, bâtiments, infrastructures). Cette analyse se base sur la quantification des aléas faite à l'étape précédente.
\end{enumerate}}

\add{Les méthodes utilisées dépendent de la taille du site d'étude, des objectifs (e.g. cartes informatives ou règlementaires) et des données disponibles. }Nous nous intéressons ici plus particulièrement à \add{l'étape de caractérisation de la propagation, pour des analyses d'aléas et/ou de risques à l'échelle du site, où les zones sources sont bien identifiées. Nous considérons par ailleurs seulement }la propagation des écoulements gravitaires, c'est à dire \rem{des glissements de terrain}\add{des mouvements de matériaux solides (terre/débris/roche/glace) et/ou liquides sous l'effet de la gravité et} dont la dynamique est semblable à celle d'écoulements. Les écoulements gravitaires peuvent \rem{en effet }se propager à de grandes distances de leur zone d'initiation. C'est le cas notamment des avalanches de blocs \rem{et }qui parcourent parfois plusieurs kilomètres \citep[e.g.][]{korup_longrunout_2013, lucas_frictional_2014}. De même, les écoulements chargés en eau comme les coulées de débris\add{ et les laves torrentielles} peuvent se propager sur plusieurs kilomètres ou dizaines de kilomètres \citep[e.g.][]{coussot_recognition_1996, thouret_lahars_2020}. Les vitesses associées atteignent parfois plusieurs dizaines de km~hr$^{-1}$, ce qui rend ces écoulements dangereux pour les populations, les infrastructures et les bâtiments \citep{givry_construire_2011, santi_debrisflow_2011}. 
La quantification de la propagation des écoulements gravitaires (distance de parcours, épaisseur et vitesse de l'écoulement, débit, ...) \rem{est donc très importante pour estimer}\add{permet donc de mieux estimer les potentiels dégâts, et donc} les risques associés. Cette quantification peut être réalisée de manière purement empirique à partir de bases de données d'observations. Par exemple, \citet{mitchell_rock_2019} ont établi une loi puissance reliant la distance de parcours au volume d'avalanches de blocs \citep[voir aussi, par exemple, ][]{lucas_benchmarking_2007, corominas_angle_1996}. Une approche similaire est utilisée par \citet{rickenmann_empirical_1999} et \citet{berti_dflowz_2014} pour estimer le débit, la distance de parcours et les zones inondées par des coulées de débris. Pour obtenir des résultats plus détaillés, des méthodes numériques à base physique résolvant les équations de la dynamique sont nécessaires. 


Dans cette perspective, les modèles d'écoulement en couche mince\rem{s} sont de plus en plus utilisés. Ils simulent l'épaisseur de l'écoulement et la vitesse moyennée sur l'épaisseur. Ils sont ainsi plus faciles à mettre en oeuvre que des modèles 3D qui modélisent la dynamique de chaque particule solide et/ou de chaque volume élémentaire de fluide. Ces modèles 3D sont en effet coûteux en temps de calcul, nécessitent souvent la calibration  de nombreux paramètres, et ne modélisent qu'un ou deux des processus physiques contrôlant la dynamique des écoulements gravitaires \citep{andreotti_granular_2013, delannay_granular_2017}. A l'inverse, les paramètres des modèles d'écoulement en couche mince\rem{s} sont généralement empirique\add{s}, mais moins nombreux. Cela simplifie leur calibration et l'analyse de sensibilité: dans le cas le plus simple, la mobilité de l'écoulement peut n'être contrôlée que par un seul paramètre (par exemple, un coefficient de friction basale).

Dans cette revue de la littérature, nous donnons les principales rhéologies utilisées dans les modèles \add{monophasiques }d'écoulement\add{ en} couche mince \rem{monophasiques }(i.e., les matériaux s'écoulant sont supposés homogènes), sans prise en compte de l'érosion. Nous montrons ensuite comment un modèle d'écoulement en couche mince, SHALTOP, peut être utilisé pour une étude d'aléa, en prenant l'exemple de la rivière du Prêcheur, en Martinique.
Nous discutons enfin de plusieurs axes de recherches actuels visant à améliorer les modèles existants, et à développer leur utilisation pour l'analyse d'aléas et de risques.

\section{Les modèles d'écoulement en couche mince}\label{sec:shaltop}

\subsection{Les principales rhéologies de modèles mono\rem{-}phasiques: approche historique}

Le principe de dérivation des équations d'écoulement en couche mince est d'intégrer les équations locales de conservation de la masse et des moments, perpendiculairement à la topographie, sur l'épaisseur \add{(supposée petite) }de l'écoulement. Les équations locales sont données par:
\begin{linenomath*}
\begin{align}
    \partial_t\vec{U}+(\vec{U}\cdot\nabla_{\vec{X}})\vec{U}&=-\vec{g}+\nabla\cdot\sigma, \\
\nabla_{\vec{X}}\cdot\vec{U}&=0,
\end{align}
\end{linenomath*}
où $\vec{U}(\vec{X})$ est le champ de vitesse en 3 dimensions, $-\vec{g}$ est la gravité et $\sigma$ est le tenseur des contraintes\rem{ (normalisé par la masse volumique)}. En supposant notamment que l'épaisseur de l'écoulement est faible par rapport à son étendue spatiale, et que la vitesse moyennée sur l'épaisseur est parallèle à la topographie, l'intégration des équations permet de réduire la dimension du problème.
Cette approche a été initialement développée par \citet{barre_theorie_1871}, puis \add{a été adaptée} pour des applications hydrauliques \citep[e.g.][]{dressler_new_1978}. La méthode a ensuite été \rem{adaptée}\add{reprise} par \citet{savage_motion_1989} pour modéliser des écoulements gravitaires secs. Depuis les années 2000, \rem{ils}\add{les modèles d'écoulement en couche mince} sont de plus en plus utilisés pour des applications opérationnelles \citep[e.g.][]{mcdougall_2014_2017, pastor_review_2018}. Plusieurs outils numériques commerciaux et académiques existent, comme RAMMS \citep{christen_ramms_2010}, DAN3D \citep{mcdougall_model_2004}, FLO-2D \citep{obrien_twodimensional_1993, jakob_debrisflow_2013}, r-avaflow \citep{mergili_avaflow_2017}, Volcflow \citep{kelfoun_numerical_2005} ou SHALTOP \citep{bouchut_new_2003, bouchut_gravity_2004, mangeney_numerical_2007}. 

Nous donnons ici les équations pour des écoulements homogènes, sans érosion\add{ basale}. Ces équations sont en effet les plus simples à utiliser pour des applications pratiques: les modèles plus complexes nécessitent\rem{et} des informations parfois compliquées à obtenir en pratique comme la fraction liquide ou les épaisseurs érodables. Une étape clé de l'obtention des équations est l'écriture des équations locales de conservation dans un référentiel lié à la topographie, l'intégration devant en effet se faire perpendiculairement à la topographie. Le changement de référentiel est particulièrement complexe sur des topographies 2D quelconques $Z=b(X, Y)$ à cause des dérivées spatiales de deuxième ordre \citep{bouchut_gravity_2004, luca_hierarchy_2009}. Une dérivation correcte est toutefois nécessaire pour bien décrire les effets de la courbure de la topographie, notamment pour les écoulement\add{s} rapides \citep{peruzzetto_topography_2021}. 

\begin{figure}
    \centering
    \includegraphics[width=0.8\textwidth]{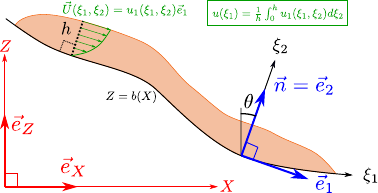}
    \caption{Notations pour les équations d'écoulement en couche mince sur une topographie 1D $Z=b(X)$. \textit{Notations for thin-layer equations on a 1D topography $Z=b(X)$}}\label{fig:notations}
\end{figure}

Par simplicité, nous donnerons seulement les équations d'écoulement en couche mince pour des topographies 1D, avec $Z=b(X)$ (Figure~\ref{fig:notations}). Considérant un système de coordonnées curvilignes associé\rem{e} $(\xi_1, \xi_2)$ ($\xi_1$ le long de la topographie, $\xi_2$ dans la direction normale à la topographie), les équations d'écoulement en couche mince ont la forme générale suivante \citep{savage_dynamics_1991, peruzzetto_numerical_2021}:
\begin{linenomath*}
\begin{align}
	\partial_th+\frac{\partial hu}{\partial{\xi_1}}&=0,\label{eq:SH_mass}\\
	\partial_t(hu)+\frac{\partial}{\partial{\xi_1}}(\alpha hu^2)+\frac{\partial}{\partial{\xi_1}}(\frac{1}{2}kgh^2\cos(\theta))&=gh\sin(\theta)-S,\label{eq:SH_mom}
\end{align}
\end{linenomath*}
avec $h$ l'épaisseur de l'écoulement, $u$ la vitesse moyennée sur l'épaisseur de l'écoulement, $\theta$ la pente de la topographie\add{ (positive pour $db/dX<0$)}, $g$ le coefficient de gravité, $\alpha$ un facteur de forme associé à des profils de vitesse non linéaire\add{s} (on prend en général $\alpha=1$), et $k$ un coefficient introduit pour prendre en compte la friction interne. Le terme source $S$ intègre les effets dissipatifs d'énergie qui ralentissement l'écoulement, par exemple par friction, turbulence et/ou viscosité. Son expression dépend donc de la rhéologie choisie, comme expliqué dans les paragraphes suivants.

\subsubsection{Modèle hydrostatique avec friction basale}\label{sec:sh_hydro}

La méthode la plus simple pour obtenir les équations d'écoulement en couche mince est de considérer que\add{ dans le référentiel lié à la topographie $(\xi_1, \xi_2)$ (voir Figure~\ref{fig:notations})} le tenseur des contraintes est donné (toujours pour un écoulement sur une topographie $Z=b(X)$) par:
\begin{linenomath*}
\begin{equation}\label{eq:sigma_1D}
 \sigma=\begin{pmatrix}
         -p &\sigma_{12}\\
         \sigma_{21} & -p
        \end{pmatrix},
\end{equation}
\end{linenomath*}
avec $\sigma_{12}=\sigma_{21}$\add{ et $p$ la pression hydrostatique}. On peut alors introduire une condition de friction à la base de l'écoulement (rhéol\add{o}gie de Coulomb):
\begin{linenomath*}
 \begin{equation}\label{eq:stress_bed_1}
    \sigma\vec{n}-\left(\vec{n}\cdot\sigma\vec{n}\right)\vec{n}=-\mu\frac{\vec{U}}{\|\vec{U}\|}(-\vec{n}\cdot\sigma\vec{n})_+\text{, à la base de l'écoulement.}
\end{equation}
\end{linenomath*}
$\mu=\tan(\delta)$ est le coefficient de friction basale, et $\delta$ l'angle de friction basale associé.\add{ Au premier ordre (en négligeant les effets d'inertie et de gradients d'épaisseurs), l'écoulement accélère s'il se propage sur des pentes $\tan(\theta)>\mu$, et ralentit et s'arrête autrement}. L'intégration des équations de conservation de la masse et des moments donne alors \citep{savage_dynamics_1991, peruzzetto_numerical_2021}:
\begin{linenomath*}
\begin{equation}\label{eq:SH_hydro}
  \partial_t(hu)+\frac{\partial}{\partial{\xi_1}}(hu^2)+\frac{\partial}{\partial{\xi_1}}(\frac{1}{2}gh^2\cos(\theta))=gh\sin(\theta)-h\mu\frac{u}{|u|}(g\cos(\theta)+\gamma u^2),
 \end{equation}
 \end{linenomath*}
où $\gamma=1/R$ est la courbure de la topographie ($R$ est le rayon de courbure). 
La généralisation de ces équations à des topographies 2D $Z=b(X,Y)$ n'est pas simple. Pour conserver le formalisme mathématique tout en arrivant à une expression similaire à \eqref{eq:SH_hydro}, \citet{bouchut_gravity_2004} choisissent
\begin{linenomath*}
\begin{equation}\label{eq:sigma}
    \sigma=\sigma'-pI_3,
\end{equation} 
\end{linenomath*}
avec\add{ $I_3$ la matrice identité et}
\begin{linenomath*}
 \begin{equation}\label{eq:visc_stress_1}
    \sigma'=\nu\left(\nabla_{\vec{X}}\vec{U}+(\nabla_{\vec{X}}\vec{U})^t\right),
\end{equation} 
\end{linenomath*}
où $\nu$ est la viscosité \rem{cinématique}\add{dynamique}, qui est supposée très petite e\rem{s}t dispara\rem{i}\add{î}t dans les équations finales. Cela revient à considérer des écoulements laminaires pour lesquels les termes dominant de $\sigma'$ sont les dérivées de la vitesse dans la direction normale à la topographie. Cela permet d'introduire formellement la condition à la base de l'écoulement \eqref{eq:stress_bed_1}, ce qui ne serait possible si on avait simplement $\sigma=-pI_3$.

\add{En réécrivant l'équation \eqref{eq:SH_hydro} dans le référentiel cartésien $(X, Z)$ et en prenant en compte les approximations faites pour obtenir l'équation \eqref{eq:SH_mom} \citep[pour plus de détails, voir][]{bouchut_gravity_2004,bouchut_bed_2022}, l'équilibre est obtenu pour:}
\begin{linenomath*}
\begin{equation}
	u=0 \text{ et } |\partial_X\left(\cos(\theta)h+Z\right)|\leq\mu=\tan(\delta)
\end{equation}
\end{linenomath*}
\add{Autrement dit, la masse reste stable tant que l'inclinaison de sa surface libre reste inférieure à $\delta$. Introduit formellement comme un coefficient de friction basale, $\mu$ traduit donc également la résistance interne des matériaux. En pratique, il est compliqué de relier $\mu$ à des propriétés géotechniques des matériaux s'écoulant, d'autant plus que ces propriétés peuvent varier lors de la propagation (par exemple par fracturation, incorporation d'eau ou variation de la nature de la surface de propagation). Ainsi, la modélisation d'écoulements gravitaires à l'échelle du terrain nécessite une étape de calibration: $\mu$ est déterminé empiriquement, et traduit tous les processus de dissipation d'énergie au sein de l'écoulement, par friction basale mais aussi, notamment, par friction/collision entre grains solides et/ou par viscosité.}

\rem{Cette approche simplifiée ne permet pas de relier formellement $\mu$ à des propriétés physiques des matériaux s'écoulant. Le coefficient de friction $\mu$ est donc empirique, et doit être choisi par calibration en reproduisant des expériences en laboratoire\rem{s} ou des écoulements gravitaires à l'échelle du terrain. Dans le cas le plus simple, $\mu$ est constant. }Il faut par exemple $\mu=\tan(2\degree)=0.03$ pour modéliser des coulées de débris\add{ particulièrement} mobiles \citep{frimberger_modelling_2021, peruzzetto_simplified_2022}. La modélisation d'avalanches rocheuses et d'avalanches de débris nécessite un coefficient de friction d'autant plus faible que le volume mobile est important \citep[typiquement, de $\mu=\tan(30\degree)=0.58$ pour un volume de 10$^3$~m$^3$, à $\mu=\tan(10\degree)=0.18$ pour un volume de $10^9$~m$^3$, ][]{lucas_frictional_2014}. Plusieurs études montrent néanmoins qu'un coefficient de friction variable, fonction de la vitesse et de l'épaisseur de l'écoulement, permet de mieux reproduire les observations. \citet{lucas_frictional_2014} proposent ainsi une loi empirique de diminution du coefficient de friction avec la vitesse, pour reproduire la mobilité importante des grands glissements de terrain:
\begin{linenomath*}
\begin{align}
    \mu&=\frac{\mu_0-\mu_w}{|u|/u_w}+\mu_w \text{ si } |u|>u_w,\\
    \mu&=\mu_0 \text{ sinon,}
\end{align}
\end{linenomath*}
\add{avec $u$ la vitesse de l'écoulement, $u_w$ une vitesse seuil déterminée empiriquement, et $\mu_0$ et $\mu_w$ deux coefficients de friction atteints respectivement pour $|u|\leq u_w$ et $|u|\rightarrow +\infty$. }\citet{lucas_frictional_2014} montrent que $\mu_0=\tan(40\degree)=0.84$, $\mu_w=\tan(6\degree)=0.11$ et $u_w=4.1$~m~s$^{-1}$ permettent de reproduire \add{la propagation d'écoulements gravitaires}\rem{des glissements de terrain} notamment sur Terre et sur Mars.

A l'inverse, une augmentation du coefficient de friction pour des vitesses importantes et/ou de faibles épaisseurs permet de simuler la chenalisation d'écoulements granulaires secs \citep{mangeney_numerical_2007, edwards_formation_2017}. Cette approche a été développée suite à des expériences de laboratoire \citep{pouliquen_friction_2002}, et a été ensuite théorisée avec la rhéologie $\mu(I)$ qui modélise la friction à l'intérieur d'écoulements granulaires (voir Section~\ref{sec:mu(I)}\add{ pour les détails de cette rhéologie}).

Mais, avant le développement de la rhéologie $\mu(I)$, une approche géotechnique a d'abord été utilisée\add{ pour distinguer la friction se produisant à l'intérieur de l'écoulement de la friction à la base de l'écoulement, au contact de la topographie}. 

\subsubsection{La friction interne: une approche géotechnique}

Dans leurs premiers travaux, \citet{savage_motion_1989} et \citet{savage_dynamics_1991} ont déduit les équations \eqref{eq:SH_mass} et \eqref{eq:SH_mom} pour des écoulements granulaires secs, et ont ainsi introduit une rhéologie frictionnelle à laquelle ils associent deux angles de friction: l'angle de friction interne $\Phi$ et l'angle de friction basal $\delta$.

Dans leurs calculs, \citet{savage_dynamics_1991} reprennent ainsi la condition de friction basale, comme dans le paragraphe précédent. Ils supposent \rem{ainsi}\add{aussi} que, pendant l'écoulement, le critère de rupture de Mohr-Coulomb est constamment vérifié. En chaque point du fluide, il existe alors un élément planaire pour lequel les contraintes normales $N$ et tangentielles $T$ vérifient:
\begin{linenomath*}
\begin{equation}\label{eq:fric_interne}
	T = \tan(\Phi) N.
\end{equation}
\end{linenomath*}
L'hypothèse de friction basale implique que les contraintes tangentielles ($T_b$) et normales ($N_b$) à la topographie vérifient:
\begin{linenomath*}
\begin{equation}\label{eq:fric_basale}
	T_b = -\frac{u}{|u|} \tan(\delta) N_b,
\end{equation}
\end{linenomath*}
où le signe dépend de la direction de la vitesse. Les relations \eqref{eq:fric_interne} et \eqref{eq:fric_basale} étant vérifiées toutes les deux à la base \rem{des }\add{de l'}écoulement, l'utilisation des diagrammes de Mohr-Coulomb permet d'identifier\add{ géométriquement} deux états de contraintes possibles: un état "passif" lors de la contraction de l'écoulement, et un état "actif" lors de sa dilatation. Le coefficient $k$ de l'équation \eqref{eq:SH_mom} peut prendre alors deux valeurs \citep{savage_dynamics_1991}: 
\begin{linenomath*}
\begin{align}
	k_{act}=2\frac{1-\sqrt{1-\left(1+\tan^2(\delta)\right)\cos^2(\phi)}}{\cos^2(\phi)}-1 \text{ quand } \partial_{\xi_1}u>0 \label{eq:kact}\\
	k_{pass}=2\frac{1+\sqrt{1-\left(1+\tan^2(\delta)\right)\cos^2(\phi)}}{\cos^2(\phi)}-1 \text{ quand } \partial_{\xi_1}u<0 \label{eq:kpass}
\end{align}
\end{linenomath*}
Ce sont les coefficients de poussées, classiquement utilisés en mécanique des sols. Cette approche permet de reproduire en partie la géométrie de dépôts expérimentaux \citep[e.g. avec $\delta=30\degree$ et $\Phi=40\degree$, ][]{ gray_gravitydriven_1999, pirulli_effect_2007}. L'extension des équations à des topographies quelconques $Z=b(X,Y)$ est toutefois difficile car l'utilisation des cercles de Mohr ne suffit plus. Dans ces situations, des simplifications sont nécessaires pour avoir une expression de la friction interne invariante par rotation \citep{christen_ramms_2010,kelfoun_numerical_2005}, ou il faut au contraire déterminer toutes les composantes du tenseur de contraintes, ce qui est beaucoup plus complexe \citep{denlinger_granular_2004}.\add{ Par ailleurs, l'utilisation d'un critère de stabilité statique (angle de friction interne $\Phi$) pour décrire un comportement dynamique peut être discuté, d'autant plus qu'il implique dans les équations \eqref{eq:kact} et \eqref{eq:kpass} que $\Phi$ soit supérieur à l'angle de friction basale $\delta$.}

La rhéologie $\mu(I)$ \rem{propose une approche plus rigoureuse}\add{, présentée dans le paragraphe suivant, propose une approche plus générale avec un coefficient de friction interne variable}. 

\subsubsection{La rhéologie \texorpdfstring{$\mu(I)$}{mu(I)} pour des écoulements granulaires}\label{sec:mu(I)}

Depuis le début des années 2000, la rhéologie de $\mu(I)$ est de plus en plus utilisée pour étudier les écoulements granulaire\add{s}. Grâce à des arguments dimensionnels et des simulations numériques, \citet{gdrmidi_dense_2004} et \citet{jop_constitutive_2006} ont montré que pour des écoulements 1D (i.e. sur des topographies $Z=b(X)$) en régime permanent, 
\begin{linenomath*}
\begin{align}
    T&=\mu(I)N, \label{eq:muI_1D}\\
    I&=\frac{\dot{\gamma}d}{\sqrt{p/\rho^*}},\label{eq:I_1D}
\end{align}
\end{linenomath*}
où $N$ \rem{est la contrainte tangentielle, $p$ la pression}\add{et $T$ sont les contraintes normales et tangentielles à la direction de l'écoulement, $p$ la pression}, $I$ le nombre d'inertie, $\dot{\gamma}$ le taux de cisaillement \rem{et }\add{, }$d$ le diamètre des grains\add{, $\rho^*$ la masse volumique des grains et $\mu(I)$ un coefficient de friction variable, fonction du nombre d'inertie $I$}.\add{ Le nombre d'inertie $I$, sans dimension, est le rapport du temps caractéristique de réarrangement des grains, sur le temps caractéristique de la déformation de l'écoulement granulaire \citep{gdrmidi_dense_2004}. Des valeurs élevées de $I$ correspondent donc à des écoulements dominés par les déformations macroscopiques par cisaillement.} L'équation \eqref{eq:muI_1D} peut \rem{ainsi }être vue comme une généralisation de l'équation \eqref{eq:fric_interne}\add{, avec un coefficient de friction $\mu(I)$ dépendant de la dynamique de l'écoulement}. Une forme tensorielle des équations \eqref{eq:muI_1D} et \eqref{eq:I_1D} \rem{ont}\add{a} été donnée\rem{s} par \citet{jop_constitutive_2006} et \citet{gray_depthaveraged_2014}\add{ pour exprimer le déviateur $\sigma'$ du tenseur de contraintes (voir équation~\eqref{eq:sigma})}:
\begin{linenomath*}
\begin{align}
    \sigma'&=\frac{\mu(I)p}{\|D\|}D, \label{eq:muI_2D}\\
    I&=\frac{2\|D\|d}{\sqrt{p/\rho^*}}. \label{eq:I_2D}
\end{align}
\end{linenomath*}
En notant $\vec{U}$ le champ de vitesse\add{s} en coordonnées cartésiennes, $D=\frac{1}{2}(\nabla_{\vec{X}}\vec{U}+(\nabla_{\vec{X}}\vec{U})^t)$ et $\|D\|=\sqrt{\frac{1}{2}tr(D^2)}$. Ces équations sont cohérentes avec une friction solide de Coulomb, car lorsque $I\rightarrow 0$, l'écoulement \rem{s'arrête}\add{se produit} seulement si 
\begin{linenomath*}
\begin{equation}
    \|\sigma'\|>\mu_sp,
\end{equation} 
\end{linenomath*}
où $\mu_s$ est un coefficient de friction donné \citep{jop_constitutive_2006}. 

La rhéologie de $\mu(I)$ a été utilisée pour dériver les équations d'écoulement en couche mince sur des topographies 1D par \citet{gray_depthaveraged_2014}, en supposant des écoulement\add{s} en régime permanent\add{ (ce qui implique un profil de vitesse de Bagnold)}: 
\begin{linenomath*}
\begin{equation}\label{eq:SH_muI}
 \partial_t(hu)+\frac{\partial}{\partial{\xi_1}}(hu^2)+\frac{\partial}{\partial{\xi_1}}(\frac{1}{2}gh^2\cos(\theta))=gh\sin(\theta)-S,
\end{equation}
\end{linenomath*}
avec
\begin{linenomath*}
\begin{equation}\label{eq:SH_muI2}
 S =\mu(I)gh\cos(\theta)\frac{u}{|u|} - \frac{\partial}{\partial \xi_1}\Bigl(\nu h^{\frac{3}{2}}\frac{\partial u}{\partial \xi_1}\Bigr).
\end{equation}
\end{linenomath*}
$S$ fait apparaître un terme frictionnel avec un coefficient \rem{de friction }de friction $\mu(I)$ dépendant du nombre d'inertie, et un terme "visqueux" introduisant le paramètre $\nu$ (qui peut être formellement relié à $I$ et à la pente locale de la topographie). L'étude expérimentale\rem{s} d'écoulements granulaires donne \rem{pour $\mu(I)$ }\citep{jop_constitutive_2006}:
\begin{linenomath*}
\begin{equation}\label{eq:muI_approx}
 \mu(I)=\mu_1+(\mu_2-\mu_1)\frac{1}{\frac{I_0}{I}+1},
\end{equation}
\end{linenomath*}
où $I_0$\add{, $\mu_1$ et $\mu_2$ sont des constantes et où la valeur du nombre d'inertie intégrée sur l'épaisseur de l'écoulement est:}\rem{est une constante}
\begin{equation}\label{eq:Ishaltop}
	I=\frac{5du}{2h\sqrt{gh\Phi\cos(\theta)}},
\end{equation}
\add{avec $\Phi$ la fraction solide volumique. }\citet{ionescu_viscoplastic_2015} utilisent par exemple $\mu_1=\tan(25.5\degree)=0.48$, $\mu_2=\tan(36\degree)=0.73$ et $I_0=0.279$. L'équation~\eqref{eq:muI_approx} n'est néanmoins valide que pour des écoulements permanents (plus précisément, pour des nombres de Froude élevés). Elle doit donc être adaptée pour les faibles vitesses \citep[e.g.][]{pouliquen_friction_2002, edwards_formation_2017}. 

\add{Notons que pour obtenir les équations \eqref{eq:SH_muI} et \eqref{eq:SH_muI2}, \citet{gray_depthaveraged_2014} supposent un écoulement sans glissement basal et utilisent \eqref{eq:muI_2D} : il n'y a donc pas de friction basale dans les équations initiales. Toutefois, à l'exception des termes visqueux, les équations \eqref{eq:SH_muI} et \eqref{eq:SH_muI2} peuvent aussi être obtenues en considérant la condition limite de friction basale de l'équation~\eqref{eq:stress_bed_1}, en choisissant $\mu(I)$ comme coefficient de friction basale, et sans imposer de contrainte sur la vitesse basale. Le coefficient $\mu(I)$ peut donc à la fois être vu comme un coefficient de friction interne dynamique, et un coefficient de friction basale. Comme dans le cas d'un coefficient de friction basale constant (voir section~\ref{sec:sh_hydro}), sa valeur reste donc empirique. Les limitations de la rhéologie $\mu(I)$ sont évoquées plus en détail par \citet{delannay_granular_2017}: ces limites incluent, entre autre, le fait que la rhéologie $\mu(I)$ ne soit pas reliée par une analyse théorique aux interactions à l'échelle des grains. Par ailleurs, les formulations existantes de la rhéologie ne sont pas mathématiquement bien posées. Cela peut conduire à des instabilités numériques dans les simulations pour des valeurs de $I$ trop élevées ou trop faibles. Enfin, l'extension des équations d'écoulement en couche mince avec la rhéologie $\mu(I)$ à des topographies 2D $Z=b(X,Y)$ n'a été faite que pour le cas d'un plan incliné $Z(X, Y)=\tan(\theta)X$ \citep{baker_twodimensional_2016}.}

\rem{L'extension de\rem{s} ces équations sur une topographie plane 2D $Z(X, Y)=\tan(\theta)X$ a été faite par \citet{baker_twodimensional_2016}, mais aucune généralisation sur des topographies quelconques n'existe. Toutefois}\add{Malgré ces limites}, l'utilisation empirique de la rhéologie $\mu(I)$ \add{avec la formulation présentée plus haut }sur des topographies quelconques donne de meilleurs résultats qu'un coefficient de friction constant. La rhéologie $\mu(I)$ permet en effet de mieux reproduire les dépôts de glissements de terrain \add{aériens} \citep[e.g.][]{guimpier_dynamics_2021}\add{ et sous-marins \citep{brunet_numerical_2017}}, l'auto-chenalisation des écoulements \citep{mangeney_numerical_2007, mangold_sinuous_2010}, et l'augmentation de la distance de parcours lors de la présence d'un lit érodable \citep{fernandez-nieto_multilayer_2016}.

\subsubsection{Modèles hydrauliques et visco-plastiques}

Des approches hydrauliques sont aussi parfois utilisées, en particulier pour la modélisation des coulées de débris. Par exemple, la rhéologie associant les lois de Darcy-Weisbach et Manning donne \add{pour $S$ dans l'équation~\eqref{eq:SH_mom}} \citep{chow_openchannel_1959, obrien_twodimensional_1993, jakob_debrisflow_2013}:
\begin{linenomath*}
	\begin{equation}\label{eq:T_darcy}
		S=gn^2\frac{u^2}{h^{1/3}},
	\end{equation}
\end{linenomath*}
où $n$ \add{en s~m$^{-1/3}$ }est le coefficient de Manning associé à la rugosité de la topographie \citep[e.g. $n=0.02$ à $n=0.1$ dans ][]{jakob_debrisflow_2013}. Cette équation a été déduite empiriquement pour des écoulements permanents dans des chenaux ouverts. 

La modélisation des écoulements chargés en eau avec une fraction solide argileuse (comme certaines laves torrentielles) nécessite l'utilisation de rhéologies visco-plastiques \citep[e.g.][]{coussot_rheology_1993, laigle_numerical_1997}. La loi de Bingham décrit ainsi des écoulement\add{s} visqueux à seuil. Pour un écoulement sur une topographie 1D $Z=b(X)$, elle donne l'expression de la contrainte de cisaillement \add{(toujours dans le référentiel $(\xi_1, \xi_2)$ lié à la topographie)}: 
\begin{linenomath*}
\begin{align}
    \sigma_{12}&=\tau_y+\nu \frac{\partial u_1}{\partial \xi_3} \text{  pour  } \left|\frac{\partial u_1}{\partial \xi_3}\right|\ge 0 \label{eq:bingham1},\\
    \sigma_{12} &\leq \tau_y \text{  pour  } \left|\frac{\partial u_1}{\partial \xi_3}\right|=0, \label{eq:bingham2}
\end{align}
\end{linenomath*}
où $\tau_y$ est la contrainte de rupture ("yield stress" en anglais)\rem{,}\add{ et} $\nu$ est la viscosité \add{dynamique}\rem{cinématique et $\rho$ la masse volumique}. Dans la partie supérieure de l'écoulement, la contrainte de cisaillement est inférieure à $\tau_y$ et la vitesse est constante. Dans la partie inférieure, la vitesse présente un profil parabolique. Suivant les travaux de \citet{pastor_simple_2004}, la contrainte basale \rem{$T$ }est alors liée à la vitesse moyenne de la colonne de fluide par une équation du troisième degré\add{ \citep[voir aussi ][]{peruzzetto_numerical_2021}}. Une solution simplifiée\add{ pour $S$} est donnée par:
\begin{linenomath*}
\begin{equation}
    S=\frac{3}{2}\frac{\tau_y}{\rho}+3\frac{\nu}{\rho}\frac{u}{h},
\end{equation} 
\end{linenomath*}
\add{avec $\rho$ la masse volumique. }La contrainte basale augmentant pour des vitesses élevées et des petites épaisseurs (comme pour la rhéologie $\mu(I)$), la rhéologie de Bingham permet de reproduire l'auto-chenalisation et la formation de levées sur les côtés des écoulements \citep{coussot_rheology_1993}. Des formes plus complexes \rem{de}\add{que} la loi de Bingham ont été proposées. Par exemple, la loi de Herschel-Bulkley permet de modéliser des comportements non linéaires \citep[][]{laigle_numerical_1997, remaitre_flow_2005}: 
\begin{linenomath*}
 \begin{equation}
    \sigma_{12}=\tau_y+ k\left(\frac{\partial u_1}{\partial \xi_2}\right)^m,
\end{equation} 
\end{linenomath*}
où $k$ \add{est le facteur de consistance (Pa~s$^m$) et correspond à une viscosité pour $m=1$, }et $m$ \add{est l'indice de rhéofluidification, sans unité}\rem{sont des paramètres du modèles}. L'analyse rhéométrique\rem{s} d'échantillons montre que $m=1/3$ peut souvent être utilisé pour les coulées de débris visqueuses \citep{remaitre_flow_2005}. Pour une masse volumique $\rho=1800$~kg~m$^{-3}$, $\tau_y$ varie typiquement entre 30 et 1000~Pa, et $k$ entre 8 et 200 Pa~s$^{-m}$ \citep{remaitre_flow_2005}. Par l'analyse des équations et la comparaison avec des expériences, \citet{coussot_steady_1994} montre que pour des écoulements permanents sur des plans infinis, et pour $m=1/3$, on obtient:
\begin{linenomath*}
 \begin{equation}
    S=\frac{\tau_y}{\rho}\left[1+1.93\left(\frac{\tau_y}{k}\left(\frac{h}{u}\right)^{\frac{1}{3}}\right)^{-0.9}\right].
\end{equation} 
\end{linenomath*}
Si $\tau_y=0$, \citet{pastor_viscoplastic_2015} calculent:
\begin{linenomath*}
 \begin{equation}
    S=\frac{k}{\rho}\left(\frac{1+2m}{m}\right)^m\left(\frac{u}{h}\right)^m.
\end{equation} 
\end{linenomath*}
En particulier, en prenant $m=2$ et en ajoutant un terme de friction, on obtient:
\begin{linenomath*}
\begin{equation}
    S= g h \cos{\theta}\tan{\delta}+\frac{25k}{4\rho}\frac{u^2}{h^2}.
\end{equation} 
\end{linenomath*}
Cette expression peut être reliée à la loi empirique de Voellmy \citep{voellmy_uber_1955}, couramment utilisée pour modéliser des avalanches et des coulées de débris \citep{mcdougall_2014_2017}:
\begin{linenomath*}
\begin{equation}
    S= g h \cos{\theta}\tan{\delta}+ g\frac{u^2}{\xi},
\end{equation}
\end{linenomath*}
où $\xi$ est un paramètre empirique appelé le coefficient de turbulence, généralement choisi entre 100 et 500~m~s$^{-2}$\citep{zimmermann_2d_2020}. 

\begin{table}[]
\caption{Principales rhéologies pour les modèles\add{ monophasiques} d'écoulement en couche mince\rem{s} \rem{mono\rem{-}phasiques}. Les notations en gras sont les paramètres à choisir dans les simulations. Leur description est donnée\rem{s} dans le corps de l'article. Des comparaisons de simulations avec différentes rhéologies peuvent par exemple être trouvées dans \citet{mcardell_systematic_2003} ou \citet{pirulli_results_2008} \textit{Main rheologies for homogeneous thin-layer models. Bold symbols are the parameters to be chosen in simulations. Their description is given in the main body of the article. Comparisons of simulations with different rheologies can be found, for instance, in \citet{mcardell_systematic_2003} and \citet{pirulli_results_2008}}}\label{tab:rheol}
\begin{adjustbox}{width=1\textwidth}
\normalsize
\begin{tabular}{|c|c|c|c|c|}
\hline
Type de rhéologie                & Nom                  & \rem{Contrainte basale $T$}\add{Terme source $S$ dans \eqref{eq:SH_mom}}                                                                                                                       & Remarque                                                                                                                                                                                         & Référence                                                                                                                                                                                               \\ \hline
\multirow{4}{*}{Frictionnelle}   & Coulomb              & \multirow{4}{*}{$S=h\bm{\mu}(g\cos(\theta)+\gamma u^2)$}                                                                                    & $\mu$ constant                                                                                                                                                                                   & \citet{savage_dynamics_1991}                                                                                                                                                                            \\ \cline{2-2} \cline{4-5} 
                                 & Friction interne     &                                                                                                                                             & \begin{tabular}[c]{@{}c@{}}$\mu$ constant\\ Angle de friction interne $\bm{\Phi}$\\ pour calculer les coefficients\\ de pressions $k_{act/pass}$\end{tabular}                                    & \citet{savage_dynamics_1991}                                                                                                                                                                            \\ \cline{2-2} \cline{4-5} 
                                 & $\mu(I)$             &                                                                                                                                             & \begin{tabular}[c]{@{}c@{}}$\mu=\bm{\mu_1}+(\bm{\mu_2}-\bm{\mu_1})\frac{1}{\bm{I_0}/I+1}$\\ + \textbf{formule de transition} pour les\\ faibles vitesses/petites épaisseurs\end{tabular} & \begin{tabular}[c]{@{}c@{}}\citet{gdrmidi_dense_2004}\\ \citet{jop_constitutive_2006}\\ \citet{gray_depthaveraged_2014}\\ \citet{pouliquen_friction_2002}\\ \citet{edwards_formation_2017} \\ \citet{mangeney_numerical_2007}\end{tabular} \\ \cline{2-2} \cline{4-5} 
                                 & Frictional weakening &                                                                                                                                             & \begin{tabular}[c]{@{}c@{}}$\mu=\frac{\bm{\mu_0}-\bm{\mu_w}}{|u|/\bm{u_w}}+\bm{\mu_w}$ si $|u|>\bm{u_w}$\\ $\mu=\bm{\mu_w}$ sinon\end{tabular}                                                   & \citet{lucas_frictional_2014}                                                                                                                                                                           \\ \hline
Hydraulique                      & Darcy-Manning        & $S=g\bm{n^2}\frac{u^2}{h^{1/3}}$                                                                                                             & -                                                                                                                                                                                                & \begin{tabular}[c]{@{}c@{}}\citet{chow_openchannel_1959}\\ \citet{obrien_twodimensional_1993}\\ \citet{jakob_debrisflow_2013}\end{tabular}                                                              \\ \hline
\multirow{3}{*}{Visco-plastique} & Bingham              & $S=\frac{3}{2}\bm{\frac{\tau_y}{\rho}}+3\bm{\frac{\nu}{\rho}}\frac{u}{h}$                                                                                & \begin{tabular}[c]{@{}c@{}}Solution simplifiée de\\ la rhéologie de Bingham\end{tabular}                                                                                                         & \begin{tabular}[c]{@{}c@{}}\citet{pastor_simple_2004}\end{tabular}                                                                                                \\ \cline{2-5} 
                                 & Herschel-Bulkley (1) & $S=\bm{\frac{\tau_y}{\rho}}\left[1+1.93\left(\frac{\bm{\tau_y}}{\bm{k}}\left(\frac{h}{u}\right)^{\frac{1}{3}}\right)^{-0.9}\right]$ & \begin{tabular}[c]{@{}c@{}}Dérivé numériquement et\\ expérimentalement\\ pour $m=1/3$\end{tabular}                                                                                               & \begin{tabular}[c]{@{}c@{}}\citet{coussot_steady_1994}\\ \citet{laigle_numerical_1997}\end{tabular}                                                                                                    \\ \cline{2-5} 
                                 & Herschel-Bulkley (2) & $S=\frac{\bm{k}}{\bm{\rho}}\left(\frac{1+2\bm{m}}{\bm{m}}\right)^{\bm{m}}\left(\frac{u}{h}\right)^{\bm{m}}$                                            & \begin{tabular}[c]{@{}c@{}}Dérivé numériquement\\ pour $\tau_y=0$\end{tabular}                                                                                                                   & \citet{pastor_viscoplastic_2015}                                                                                                                                                                           \\ \hline
Empirique                        & Voellmy              & $S= h\bm{\mu}\left(g\cos(\theta)+\gamma u^2\right)+\frac{u^2}{\bm{\xi}}$                                                                    & Formule empirique                                                                                                                                                                                & \begin{tabular}[c]{@{}c@{}}\citet{voellmy_uber_1955}\end{tabular}                                                                                                              \\ \hline
\end{tabular}
\end{adjustbox}
\end{table}

Les différentes rhéologies présentées plus haut sont résumées dans le Tableau~\ref{tab:rheol}. Elles sont toutes empiriques, dans la mesure où les paramètres associés sont difficiles à relier aux propriétés physiques des matériaux. Les paramètres sont en effet soit fondamentalement empiriques (comme le coefficient de friction de Voellmy), soit complexes à mesurer \citep[comme la viscosité de coulées de débris;][]{sosio_field_2007}. Les paramètres des modèles d'écoulement en couche mince doivent donc être calibrés en reproduisant des événements connus, avec la difficulté d'avoir à disposition des données pour caractériser ces événements (par exemple, volumes mobilisés, vitesse\add{s} et épaisseurs des dépôts). Notons par ailleurs que, à notre connaissance, aucun modèle ne permet de combiner une description rhéologique complexe des écoulements avec une dérivation rigoureuse des équations sur des topographies quelconques. L'approche la plus générale est donnée par \citet{luca_hierarchy_2009}, mais elle-même nécessite des hypothèses sur les profils de vitesse au \rem{sien}\add{sein} de l'écoulement. Par ailleurs, les équations de \citet{luca_hierarchy_2009} n'ont pas été implémentées dans un modèle numérique.

Ainsi, il peut être préférable d'utiliser des rhéologies plus simples nécessitant la calibration de moins de paramètres, tout en utilisant des équations décrivant le plus précisément possible la géométrie de topographies complexes. C'est cette approche qui a été choisie pour le code SHALTOP. 

\subsection{SHALTOP}

\add{Le modèle }SHALTOP modélise la propagation d'écoulements homogènes, sans érosion, sur des topographies complexes \citep{bouchut_new_2003, bouchut_gravity_2004, mangeney_numerical_2007}. Il a été testé pour modéliser des expériences de laboratoire \citep{mangeney-castelnau_use_2005} et des avalanches de débris ou de roches \citep[e.g.][]{lucas_frictional_2014, moretti_numerical_2015}. Contrairement à la majorité des autres codes d'écoulement en couche mince, SHALTOP prend en compte précisément la courbure de la topographie dans les équations. \citet{peruzzetto_topography_2021} ont montré que la courbure de la topographie peut avoir une influence \rem{importante}\add{significative} sur la dynamique des écoulements rapides, en particulier pour modéliser les débordements d'écoulements chenalisés.

Pour des topographies 1D $Z=b(X)$, les équations résolues par SHALTOP sont celles données par les équations \eqref{eq:SH_mass} et \eqref{eq:SH_hydro}. Sur des topographies 2D $Z=b(X,Y)$, leur forme est plus complexe. Les équations incluent notamment le tenseur de courbure de la topographie. Par souci\rem{s} de simplicité, nous ne les donnons pas ici. Elles peuvent être trouvées dans \citet{mangeney_numerical_2007}. La dérivation précise des équations est expliquée par \citet{bouchut_gravity_2004}, et la méthodologie est expliquée plus succinctement par \citet{peruzzetto_topography_2021}.

Dans sa forme la plus simple, SHALTOP n'utilise qu'une rhéologie frictionnelle de Coulomb avec un coefficient de friction $\mu_S$ constant. Néanmoins, le code permet d'utiliser des coefficients de friction non constants, dépendant par exemple de la vitesse de l'écoulement, et d'autres rhéologies (comme celles de Bingham ou de Voellmy). Dans la section suivante, nous montrons comment SHALTOP peut être utilisé pour modéliser des coulées de débris et quantifier l'exposition d'un village à ces coulées. Nous prenons le cas d'étude de la Rivière du Prêcheur, en Martinique (Petites Antilles).

\section{Exemple d'application: les coulées de débris de la rivière du Prêcheur (Martinique, Petites Antilles)}\label{sec:site}

Nous présentons ici une étude réalisée dans le cadre d'une thèse \citep{peruzzetto_numerical_2021, peruzzetto_simplified_2022} et d'un projet d'appui aux politiques publiques \citep{peruzzetto_modelisation_2021}. L'objectif de ces travaux était d'analyser les aléas associés à des avalanches de blocs générées par \rem{l'effondrement}\add{la déstabilisation} d'un escarpement rocheux, et aux coulées de débris générées par la remobilisation dans une rivière des dépôts des \rem{effondrements}\add{avalanches de blocs}. Pour ce cas d'étude en Martinique (Petites Antilles), SHALTOP a été utilisé pour modéliser la propagation des avalanches de blocs et des coulées de débris. SHALTOP ayant déjà été utilisé à plusieurs reprises pour modéliser des avalanches de blocs ou de débris \citep[e.g.][]{lucas_benchmarking_2007, moretti_numerical_2012}, nous nous concentrons ici sur la modélisation des coulées de débris. 

Après avoir présenté le site d'étude, nous détaillons les différentes étapes de l'utilisation de SHALTOP: (i) la collecte de données de terrain\rem{s} pour caractériser les phénomènes modélisés et définir les scénarios de simulation, (ii) la calibration du modèle en reproduisant un événement historique, et (iii) la simulation directe d'événements possibles pour l'analyse de risques. Les deux premiers points sont détaillés dans \citet{peruzzetto_simplified_2022}, tandis que l'analyse de risques a été faite dans \citet{peruzzetto_modelisation_2021}.

\subsection{Présentation du site}

\begin{figure}
    \centering
    \includegraphics[width=\textwidth]{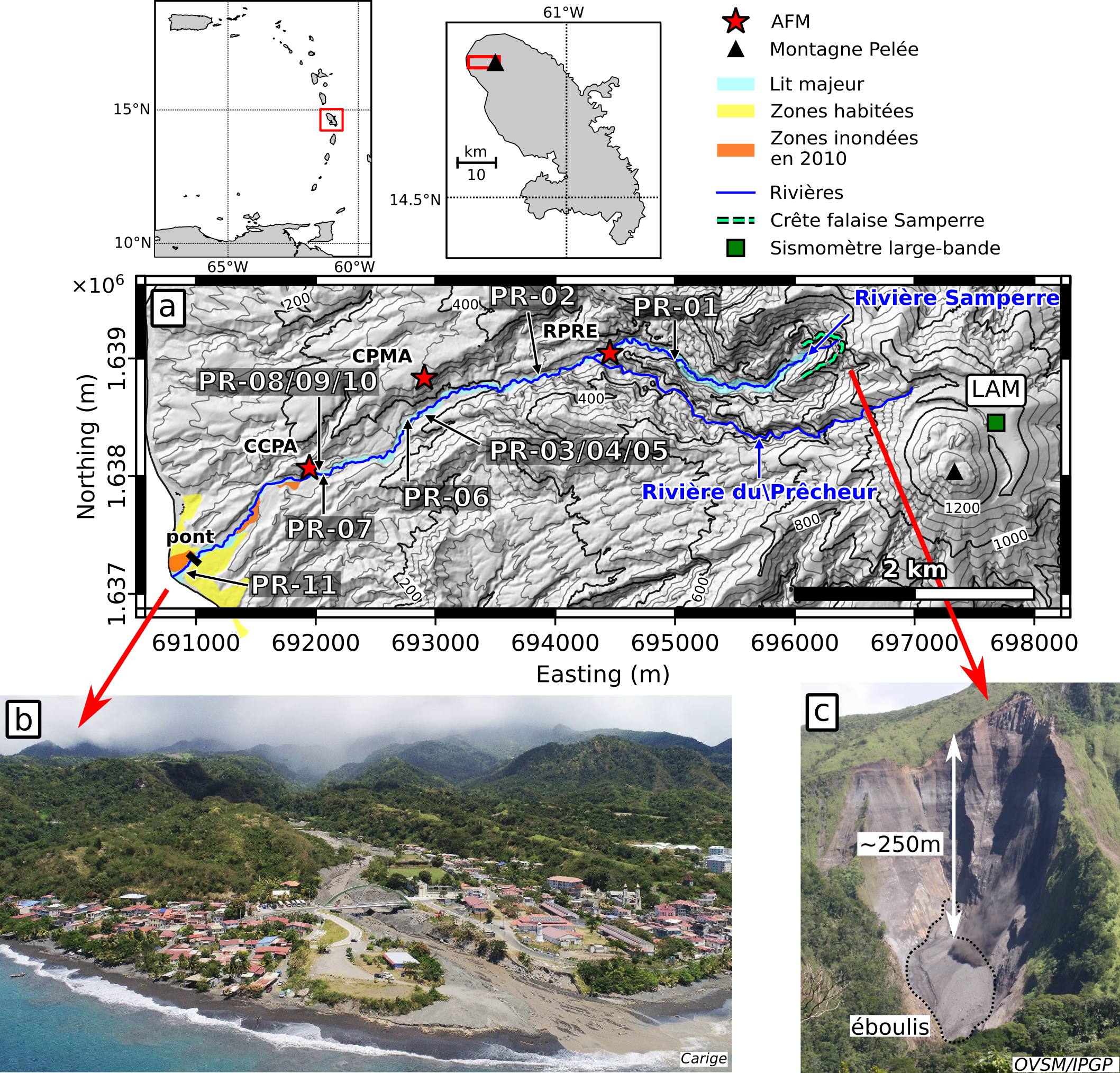}
    \caption{Le bassin versant du Prêcheur. (a)~Carte générale du site d'étude (Modèle numérique de Terrain IGN Lito3D), avec la localisation des échantillons récupérés dans le lit de la rivière (PR-XX). La localisation du site aux Antilles et en Martinique est indiquée sur les deux cartes supérieures. (b)~Vue du village du Prêcheur (30/03/2021, Carige). (c)~Vue de la Falaise Samperre (02/02/2018, OVSM/IPGP). Adapté de \citet{peruzzetto_simplified_2022}. \textit{Prêcheur river catchment. (b)~Map of the study site (Digital Elevation Model IGN Lito3D), with the location of sampling sites (PR-XX). The location of the study site in the Caribbean and in Martinique island is given by the maps above. (c)~Aerial view of the Prêcheur village (30/03/2021, Carige Company). (d)~Photograph of the Samperre cliff (02/02/2018, OVSM/IPGP). Adapted from \citep{peruzzetto_simplified_2022}.}}\label{fig:precheur_map}
\end{figure}

Le bassin versant de la Rivière du Prêcheur (Figure~\ref{fig:precheur_map}) est situé sur le versant Ouest de la Montagne Pelée en Martinique (Petites Antilles). Le principal affluent de la rivière du Prêcheur est la rivière Samperre, dont la source est située au pied\rem{s} de la Falaise Samperre. \add{7~km séparent la Falaise Samperre de l'embouchure de la rivière, avec une pente moyenne du lit comprise entre 7$\degree$ et 12$\degree$ dans la partie amont, et entre 3$\degree$ et 4$\degree$ sur les 4 derniers kilomètres. La Falaise Samperre est difficilement accessible, ce qui rend sa caractérisation géologique et géotechnique difficile. En complétant la littérature par l'analyse de photos aériennes et orthophotographies historiques, et de modèles numériques de terrain, \citet{peruzzetto_how_2022} montrent que cet escarpement est essentiellement constitué (sur 100 à 200~m) de dépôts pyroclastiques indurés de granulométries variables, contenant des blocs métriques à pluri-métriques, et probablement mis en place entre 36 et 25~ka.}

\rem{Cet escarpement}\add{La Falaise Samperre} est relativement récent\add{e}: \rem{il}\add{elle} a été formée entre 1951 et 1980 par des \rem{effondrements}\add{éboulements} successifs\add{ affectant principalement les formations pyroclastiques évoquées précédemment}. Au\rem{x} moins cinq épisodes de déstabilisations majeures se sont produits en 1980, 1997-1998, 2009-2010 et 2018 \citep{aubaud_review_2013, clouard_physical_2013, peruzzetto_how_2022}. Par exemple, \citet{clouard_physical_2013} estiment le volume effondré entre mars et mai 2010 à $2.1\times10^6$~m$^3$. Entre mai 2010 et août 2018, le volume effondré est estimé à $4.9\times10^6$~m$^3$, avec une phase paroxysmale d'effondrement début 2018 \citep{peruzzetto_simplified_2022}. La crête de l'escarpement rocheux a ainsi reculé de 250~m entre 1988 et 2018. \add{\citet{peruzzetto_how_2022} suggèrent que ces déstabilisations sont liées à la vidange d'une paléo-vallée comblée par les dépôts pyroclastiques à partir de 36~ka. La surface de la paléo-vallée favorise les écoulements d'eau souterrains et fragilise progressivement la base de l'unité pyroclastique, conduisant à sa rupture et à des déstabilisations régressives \citep[pour une discussion plus détaillée, voir][]{peruzzetto_how_2022}.}

\rem{Ces effondrements}\add{Les éboulements associés} ne menacent pas directement des zones habitées qui sont situées beaucoup plus en aval. En revanche, le réservoir de débris constitué en pied\rem{s} de falaise peut être remobilisé par l'eau sous forme de \rem{lahars, qui peuvent prendre la forme de }coulées de débris ou d'écoulements hyper-concentrés se propageant parfois jusqu'à l'embouchure de la rivière\add{, avec une ou plusieurs vagues (ou bouffées) successives}.\add{ Ces deux types d'écoulements, regroupés sous le terme générique de lahar en contexte volcanique, se différencient par la fraction solide et la manière dont elle est transportée.} Les coulées de débris présentent \rem{des}\add{une} fraction solide\rem{s} supérieure à 60\% et \rem{pas de séparation verticale entre les deux phases}\add{les phases liquides et solides sont mélangées de manière homogène}. A l'inverse, les écoulements hyper-concentrés ont une fraction solide comprise en 20\% et 60\%, et la fraction solide est transportée à la base de l'écoulement \citep{thouret_lahars_2020}. 

Les lahars, et plus spécifiquement les coulées de débris, menacent le village du Prêcheur à l'embouchure de la rivière, 7~km en aval de la Falaise Samperre. Ainsi en juin 2010, une coulée de débris a inondé la rive droite de la rivière et détruit le pont permettant d'\rem{y }accéder\add{ à cette rive}. Un nouveau pont a depuis été construit, mais des lahars importants pourraient le détruire à nouveau, isolant environ 420 personnes \citep{insee_donnees_2015}. Dans les travaux de \citet{peruzzetto_modelisation_2021} et \citet{peruzzetto_simplified_2022}, une approche conservative a été choisie: seuls les phénomènes les plus dangereux pour le village du Prêcheur sont considérés. Il s'agit des coulées de débris générées par une remobilisation \add{rapide voire }instantanée \rem{(ou tout du moins rapide) }du réservoir de matériaux situé au pied de la Falaise Samperre. 

\subsection{Caractérisation des phénomènes et rhéologie}

\rem{Avant de réaliser des simulations, il est nécessaire d'exploiter des données de terrain\rem{s} pour caractériser les événements passés et choisir les rhéologies appropriées.}

La caractérisation des événements passés a pour but d'extraire des observations quantifiées qui pourront être comparées avec les résultats des simulations lors de l'étape de calibration. Dans le cas des coulées de débris de la rivière du Prêcheur, \rem{nous avons }deux types d'observations\add{ sont disponibles}: les zones affectées par les coulées de débris, et les temps de passage des écoulements à différents points de la rivière. Les zones affectées par les débordements au niveau du village du Prêcheur sont identifiées par des photos aériennes prises juste\rem{s} après les événements. Les temps de passage sont déduits des enregistrements de géophones ("Acoustic Flow Monitoring systems", ou AFMs) installés le long de la rivière et maintenus par l'Observatoire Volcanologique et Sismologique de Martinique (OVSM). \rem{Les AFMs sont aujourd'hui largement utilisés pour détecter les lahars sur les volcans. }En Martinique ils sont utilisés pour déclencher automatiquement des alarmes dans le village du Prêcheur lorsque le débit de la rivière augmente rapidement. 

\begin{figure}
    \centering
    \includegraphics[width=\textwidth]{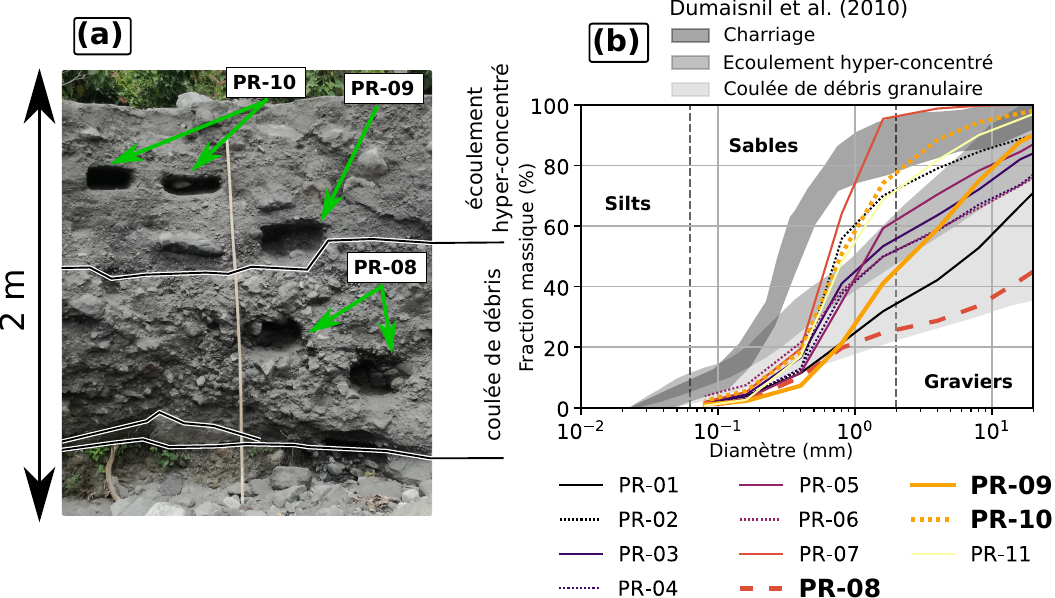}
    \caption{\rem{Analyse granulométrique}\add{Etude} des dépôts de lahars dans la Rivière du Prêcheur. (a)~Exemple de site de prélèvement, avec distinction de deux types de dépôts. (b)~Courbes granulométriques des 11 échantillons prélevés dans la rivière du Prêcheur (PR-01 à PR-11, de l'amont à l'aval, voir Figure~\ref{fig:precheur_map} pour les localisations). Les fuseaux granulométriques de \citet{dumaisnil_hydraulic_2010} sont ajoutés. Les noms en gras correspondent aux échantillons visibles en (a). Adapté de \citet{peruzzetto_simplified_2022}. \textit{Granulometric analysis and lahars deposits in the Prêcheur River. (a)~Example of sampling site, with the di\add{s}tinction \rem{of}{between} two deposits types. (b)~Granulometric curves of the 11 samples taken in the Prêcheur River (PR-01 to PR-11, from upstream to downstream, see Figure~\ref{fig:precheur_map} for locations). Granulometric ranges of \citet{dumaisnil_hydraulic_2010} are also given. Samples in bold are the samples from (a). Adapted from \citet{peruzzetto_simplified_2022}}}\label{fig:granulometry}
\end{figure}

Le choix de la rhéologie pour modéliser des coulées de débris est intrinsèquement lié à la granulométrie de la fraction solide. \citet{peruzzetto_simplified_2022} ont ainsi analysé la granulométrie de 11 échantillons de dépôts de lahars dans la rivière du Prêcheur (Figure~\ref{fig:granulometry}). L'absence d'argiles (moins de 4\% de silts et d'argiles) indique que la dynamique des écoulements est plus contrôlée par la collision et la friction entre les particules solides, que par des forces visqueuses \citep{dumaisnil_hydraulic_2010}. Ainsi une rhéologie visco-plastique n'est pas approprié\add{e} pour modéliser les coulées de débris de la rivière du Prêcheur, et des rhéologies frictionnelles sont préférables (loi de Coulomb et/ou rhéologie de Voellmy). \add{Pour caractériser plus finement les matériaux solides des coulées de débris, il faudrait échantillonner les matériaux au pied de la Falaise Samperre, mais l'accès à cette zone est trop dangereux. Par ailleurs, la comparaison des échantillons ne permet pas non plus d’identifier une variation systématique de la granulométrie en fonction de la distance à la source. Cela peut être expliqué par le fait que les dépôts de plusieurs coulées de débris ont été échantillonnés. Compte tenu du nombre de coulées de débris qui se sont produites en 2018, l’étude de terrain de 2019 n’a pas permis d’isoler les dépôts d’un même événement sur tout le linéaire de la rivière.}

\subsection{Calibration du modèle}

\citet{peruzzetto_modelisation_2021} et \citet{peruzzetto_simplified_2022} ont reproduit la coulée de débris du 19 juin 2010 pour calibrer le modèle SHALTOP. Cette coulée de débris fait suite à un\rem{e} épisode important de déstabilisations de la Falaise Samperre le 11 mai 2010, qui génère un \rem{important }réservoir de matériaux en pied\rem{s} de falaise. La coulée de débris du 19 juin 2010 a lieu entre 7h30 et 12h00 TU (Temps Universel), après une onde tropicale d'intensité non exceptionnelle \citep[11~mm pendant 1h40 avant le début du lahar;][]{aubaud_review_2013}. Sa bouffée principale a lieu entre 8h30 et 9h00 TU après 30~mm de précipitations sur une heure. La propagation de l'écoulement est alors particulièrement rapide: il parcourt 1.5~km dans la partie amont de la rivière en seulement 2 à 3 min (soit une vitesse de 30 à 45~km~hr$^{-1}$).

Cette vitesse considérée comme très rapide \citep[voir la classification des vitesses de ][]{cruden_landslide_1996} et la vidange complète du réservoir suggèrent que la coulée de débris a été initiée par une remobilisation instantanée du réservoir, peut-être par liquéfaction statique \citep[\add{c'est à dire une perte soudaine de résistance de matériaux lâches non drainés, par une mise en charge conduisant au cisaillement, à la compaction des matériaux et donc à une augmentation de la pression de pore,}][]{lalubie_lahars_2013}. Malheureusement, aucun relevé topographique ne permet de reconstruire la géométrie de ce réservoir. Ainsi, \citet{peruzzetto_simplified_2022} ont utilisé des relevés topographiques réalisés en 2018 sur lesquels un autre réservoir est identifiable, et qui est donc utilisé comme proxy du réservoir de matériaux remobilisé en juin 2010. Compte tenu des incertitudes associées à l'estimation du volume, \citet{peruzzetto_simplified_2022} ont testé plusieurs volumes pour reproduire la coulée de débris. Ils ont aussi montré qu'une initiation progressive, avec une remobilisation progressive des matériaux et non instantanée, ne permettait pas d'améliorer significativement les résultats de la calibration.

\begin{figure}
    \centering
    \includegraphics[width=\textwidth]{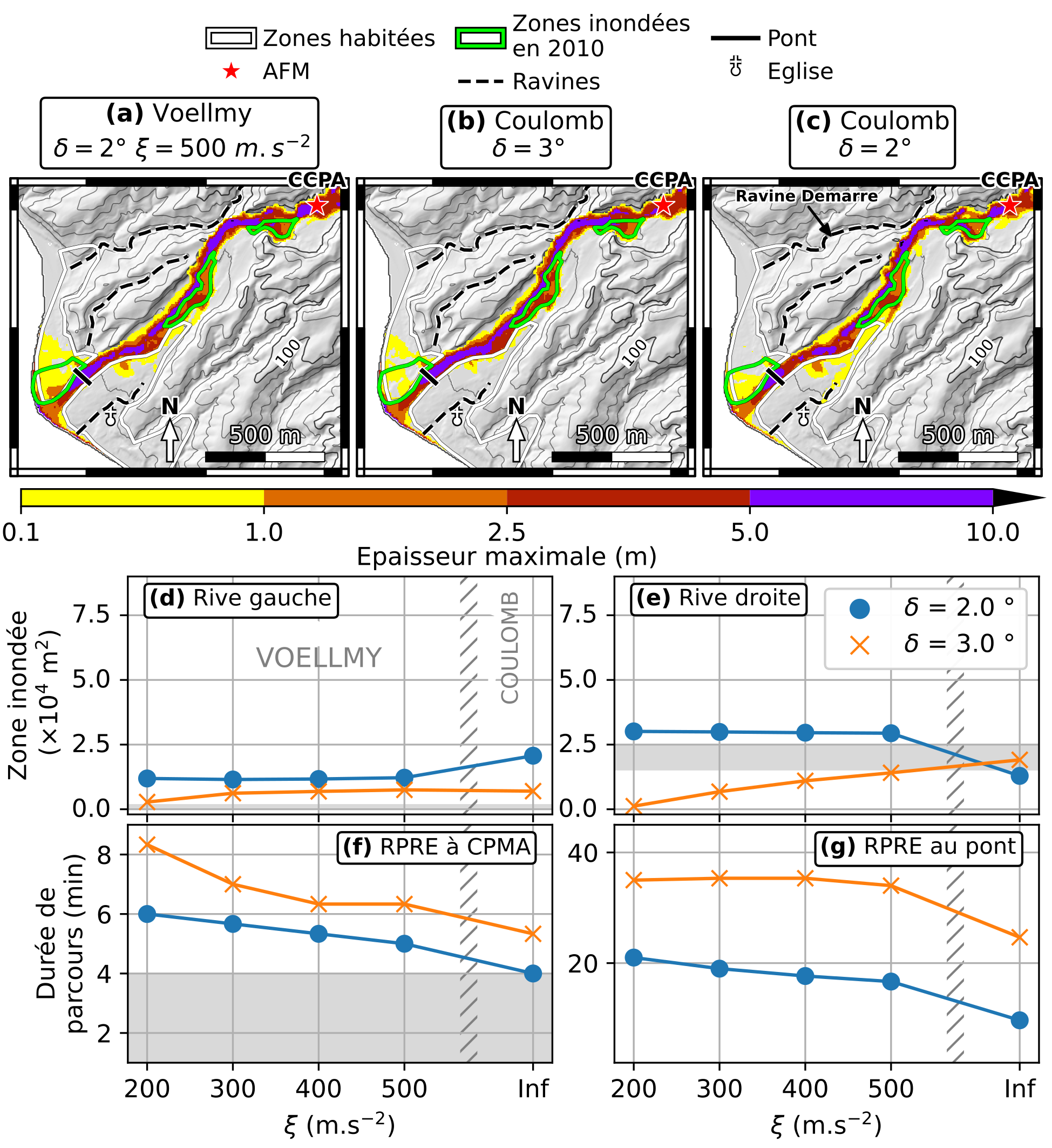}
    \caption{Résultat de la calibration de SHALTOP pour reproduire la coulée de débris du 19 juin 2010 ($0.65\times10^6$~m$^3$). (a)~Epaisseur maximale avec la rhéologie de Voellmy, $\mu_S=\tan(2\degree)$ et $\xi=500$~m~s$^{-2}$, (b)~avec la rhéologie de Coulomb et $\mu_S=\tan(3\degree)$, et (c)~avec la rhéologie de Coulomb et $\mu_S=\tan(2\degree)$. Chaque point dans (d\rem{(}\add{)}, (e), (f), et (g) est un résultat de simulation, avec le coefficient de friction donné par la couleur de la ligne et le co\add{e}fficient de turbulence donné par l'abcisse. Les points à gauche de la ligne hachurée sont les résultats avec Voellmy, ceux à droite les résultats avec \add{la rhéologie de }Coulomb. (d) et (e): surfaces inondées sur les rives gauches et droites, dans les zones habitées. (f) et (g): durée de parcours entre RPRE et CPMA (1,6~km), et RPRE et le pont (4,3~km). Les zones grisées correspondent aux observations, en prenant en compte les incertitudes. 
    \textit{Simulation results for the Jun\rem{.}\add{e} 19, 2010 debris flow ($0.65\times10^6$~m$^3$). (a)~Maximum flow thickness with the Voellmy rheology, $\mu_S=\tan(2\degree)$ and $\xi=500$~m~s$^{-2}$, (b)~with the Coulomb rheology and $\mu_S=\tan(3\degree)$, and (c)~with the Coulomb rheology and $\mu_S=\tan(2\degree)$. Each point in (d), (e), (f) and (g) is a simulation result, with friction coefficient given by line color and turbulence coefficients given by the x-coordinate. Left of hatches is for the Voellmy rheol\add{o}gy, right is for the Coulomb rheology. (d) and (e): Area flooded on the left and right riverbank, within inhabited areas. (f) and (g): Flow travel duration between RPRE and CPMA (about 1.6~km), and between RPRE and the Pr\^echeur bridge (about 4.3~km). Grey patches are observations, taking into account uncertainties.}}\label{fig:calibration}
\end{figure}

Le meilleur accord entre les observations et les résultats de simulation est ainsi obtenu pour un volume de $0.65\times 10^6$~m$^3$ remobilisé instantanément. La Figure~\ref{fig:calibration} présente les résultats des simulations avec la rhéologie de Coulomb et la rhéologie de Voellmy. Les meilleurs résultats sont obtenus avec Coulomb et $\mu=\tan(2\degree)$: les zones inondées rive gauche sont sur\rem{-}estimées, mais la zone inondée rive droite est correctement reproduite. Le temps de parcours simulé entre les stations RPRE et CPMA est de 4 minutes, ce qui est cohérent avec le temps de parcours estimé à partir des enregistrements des AFMS. 

Ainsi, le choix d'une rhéologie simple de Coulomb, semi-emprique, n'utilisant qu'un seul paramètre, permet de reproduire les caractéristiques principales d'une coulée de débris de forte intensité. C'est donc cette rhéologie qui a été utilisée pour réaliser une analyse \add{d'aléas puis }de risques.

\subsection{Modélisation directe et analyse de risques}

\begin{figure}
    \centering
    \includegraphics[width=0.8\textwidth]{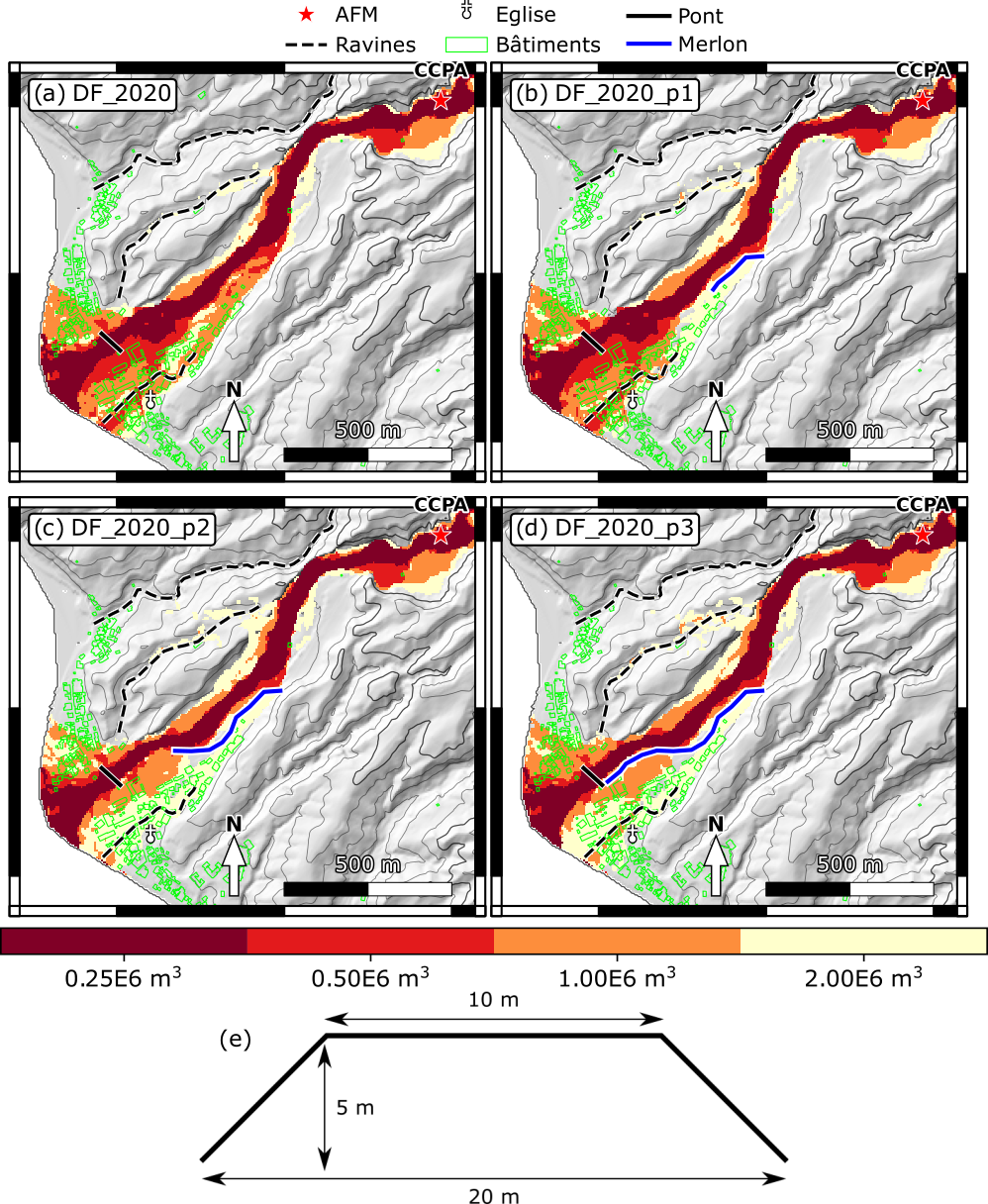}
    \caption{Etude de l'influence d'un merlon sur l'exposition du village du Prêcheur aux coulées de débris. L'échelle de couleur donne les zones impactées en fonction du volume des coulées de débris. (a)~Cas de référence, avec la topographie d'août 2018 (DEAL Martinique/Helimap). (b)~Avec un merlon de 175~m de long. (c)~Avec un merlon de 400~m de long. (c)~Avec un merlon de 575~m de long. (e) Dimensions du merlon. Adapté de \citet{peruzzetto_modelisation_2021}.
    \textit{Analysis of the influence of protection walls on the vulnerability of the Prêcheur village to debris flows. The colorscale indicates which areas are affected by debris flows, depending on their volumes. (a)~Reference case, with the topography of August 2018 (DEAL Martinique/Helimap). (b)~With a 175~m protection wall. (c)~With a 400~m protection wall. (d)~With a 575~m protection wall. (e)~Dimensions of the protection wall. Adapted from \citet{peruzzetto_modelisation_2021}.}}\label{fig:merlons}
\end{figure}

 Des coulées de débris \add{de }volumes variés (de $0.05\times10^6$~m$^3$ à $2\times10^6$~m$^3$) ont été modélisées par \citet{peruzzetto_modelisation_2021}, avec différentes géométries du lit de la rivière dans son dernier kilomètre. Plusieurs configurations sont considérées, avec un comblement, un creusement et/ou un élargissement du lit de la rivière\add{ préalablement à l'occurrence de la coulée de débris}. Des merlons de protection\rem{s} de différentes longueurs sont également simulés. Dans chaque cas, l'exposition du village du Prêcheur est quantifié\add{e} en terme de nombres de bâtiments impactés. Pour des coulées de débris de volume\add{s} supérieurs à $1.5\times10^6$~m$^3$, les différentes configurations ne changent pas l'exposition du village du Prêcheur, dans la mesure ou des débordements majeurs se produisent de toute façon. En revanche, pour des volumes inférieurs à $1.0\times10^6$, le maintien du lit de la rivière à son niveau le plus bas et l'installation d'un merlon de protection rive gauche permet de réduire de 50\% le nombre de bâtiments impactés (Figure~\ref{fig:merlons}).

Ce pourcentage dépend néanmoins des dimensions du merlon, et des volumes considérés. Les simulations numériques permettent en effet de mettre en évidence que l'augmentation de la longueur du merlon n'améliore pas systématiquement la protection du village. \citet{peruzzetto_modelisation_2021} analyse\rem{nt} ainsi l'efficacité de 3 merlons de 5 mètres de haut mis en place sur la rive gauche: un merlon court (175~m de long), un merlon intermédiaire (400~m de long) et un merlon long (575~m). Par exemple, le merlon intermédiaire protège moins le village que le merlon le plus court, pour une coulée de débris de $0.65\times 10^6$~m$^3$. En effet, l'allongement du merlon concentre la coulée dans la rivière et engendre des débordements plus importants en aval. De même, le merlon de 575~m de long engendre plus de débordements rive droite que le merlon intermédiaire. \add{Dans d'autres contextes, ces résultats pourraient être utilisés pour identifier les potentielles zones d'épandage, où les coulées de débris seraient en partie déviées. Mais pour la rivière du Prêcheur, les contraintes topographiques et l'urbanisation autour de l'embouchure de la rivière empêchent, en l'état, de définir de telles zones d'épandage. }

L'utilisation de SHALTOP permet ainsi d'analyser l'exposition du village du Prêcheur, et d'aider à dimensionner les ouvrages de protection. Ces résultats doivent néanmoins être pris avec prudence, car toute la complexité des phénomènes n'est pas modélisée. Dans la section suivante, nous identifions trois axes de recherches \rem{importants }pour améliorer l'utilisation opérationnelle\rem{s} des codes d'écoulement en couche mince.

\section{Perspectives de développement et d'utilisation des modèles d'écoulement\rem{s} en couche mince}

\subsection{Développement\add{s} rhéologiques et numériques}

 Dans la plupart des cas, les applications opérationnelles (comme celle présentée ci-dessus) considèrent des écoulements mono\rem{-}phasiques et homogènes. Des développements récents s'intéressent à la modélisation d'écoulements bi\rem{-}phasiques \citep[][]{bouchut_twophase_2016, pastor_twophase_2018, iverson_depthaveraged_2014, mergili_avaflow_2017}. Dans le cas de la rivière du Prêcheur, de tels modèles pourraient aider à mieux modéliser l'initiation des coulées de débris (avec l'intégration progressive de matériaux solides dans une phase liquide) et leur arrêt (sédimentation progressive de la fraction solide). De même, des modèles multi\rem{-}couches \citep{fernandez-nieto_multilayer_2016, fernandez-nieto_2d_2018} peuvent être utilisés pour analyser plus précisément les variations de vitesses verticales au sein d'une colonne de fluide, et tenter de modéliser des écoulements hyper-concentrés (\rem{où les phases liquides et solides présentent une séparation verticale}\add{avec une fraction solide/liquide variable sur une même verticale}). Toutefois, comme évoqué précédemment, de tels modèles peuvent être complexes à calibrer et/ou ne sont pas encore adaptés à des rhéologies complexes.
 
 Une autre piste de recherche largement étudiée depuis plusieurs années est la modélisation de l'érosion \add{basale et latérale }(voir \citet{iverson_entrainment_2015} pour une revue de la littérature). Celle-ci peut changer significativement le volume des coulées de débris: par exemple, le volume de la coulée de débris de Tsing Shan (Hong-Kong, 2000) a été multiplié par 10 entre son\rem{t} initiation et son arrêt \citep{pirulli_numerical_2012}. Toutefois, la prise en compte de l'érosion dans les équations d'écoulement en couche mince est difficile à concilier, d'un point de vue méthodologique, avec la contrainte de conservation de l'énergie \citep{bouchut_new_2008, iverson_entrainment_2015, pudasaini_mechanical_2020}. Un\add{e} possibilité est d'utiliser un modèle multi\rem{-}couche avec un\add{e} zone initialement à l'arrêt à la base de l'écoulement \citep{fernandez-nieto_multilayer_2016}, mais il n'existe pas d'approche unifiée \citep{pirulli_numerical_2012, lusso_explicit_2021}. Par ailleurs, les résultats des simulations dépendent fortement d'une connaissance experte des zones et des épaisseurs érodables, ce qui peut être compliqué à estimer en pratique. Dans les cas où l'érosion se produit essentiellement près de la zone source, l'augmentation du volume qui en découle peut être prise en compte, dans une première approximation, en faisant simplement varier le volume initial \citep{peruzzetto_simplified_2022}. Mais dans d'autres cas où l'érosion peut avoir lieu sur toute la zone de propagation (par exemple, lorsque des dépôts de coulées de débris non consolidés sont remobilisés par une nouvelle coulée de débris), la prise en compte de l'érosion devient nécessaire \citep{reid_forecasting_2016}.

\subsection{Analyse des aléas et des risques}

Actuellement en France, la cartographie des aléas pour les Plans de Prévention des Risques est faite principalement de manière experte \citep{thiery_improvement_2020}. Des méthodes empiriques quantitatives peuvent être utilisées pour quantifier de manière objective et répétable la propagation à l'échelle locale et régionale \citep[échelle 1:5 000 à 1:250 000, e.g.][]{mergili_combining_2019, guyomard_alea_2021}. Les modèles d'écoulement en couche mince sont plutôt utilisés, à l'heure actuelle, pour des études sur des sites plus petits, avec des zones sources identifiées. Quelques exemples d'utilisation des modèles d'écoulement en couche mince pour quantifier la propagation à l'échelle régionale existent toutefois. Par exemple \citet{bout_integration_2018} modélisent la rupture superficielle de versants et les coulées de débris associées dans une même simulation. Ils ne considère\add{ent} toutefois qu'un seul scénario, et ne proposent pas de méthodologie pour obtenir une carte d'aléa.

Une analyse de risques détaillée nécessite de relier l'intensité des phénomènes aux dommages des infrastructures et bâtiments.
Par exemple, pour les coulées de débris, il n'existe pas de consensus sur la définition d'un indicateur d'intensité. Un indicateur classique (car facilement observable sur le terrain) est la hauteur de l'écoulement \citep[e.g.][]{papathoma-kohle_loss_2015}. Mais d'autres indicateurs sont parfois utilisés comme la vitesse \citep{dikau_landslide_1996}, la pression d'impact \citep{scheidl_analysing_2013, givry_construire_2011}, ou un critère combiné\rem{e} sur la vitesse et l'épaisseur \citep{givry_construire_2011, hurlimann_evaluation_2008}. Dans tous les cas, l'obtention de courbes d'endommagement (degré de dommage en fonction de l'intensité du phénomène) pour un type de bâti est associé\add{e} à de fortes incertitudes car les indicateurs d'intensité sont difficiles à estimer, en particulier ceux associés à la dynamique de l'écoulement, comme la pression d'impact \citep[e.g.][]{jakob_vulnerability_2012}.
La simulation d'événements passés avec les modèles d'écoulement\rem{s} en couche\rem{s} mince\rem{s} pourrait permettre de mieux caractériser le lien entre les dommages des bâtiments et infrastructures, et l'intensité des phénomènes. En effet, les relevés de terrain\rem{s} réalisés après les événements ne donnent qu'une vision parcellaire de la dynamique de l'écoulement, alors que la simulation numérique permet d'estimer, en chaque point, l'évolution des vitesses et des épaisseurs. 

Pour obtenir des résultats précis, l'interaction entre l'écoulement et les bâtiments doit être modélisée. Cette interaction est parfois prise en compte de manière implicite en considérant les zones urbaines comme des zones poreuses \citep{sanders_integral_2008}, ou de manière explicite en délimitant les bâtiments \citep{ouro_immersed_2016}. Ces méthodes sont souvent utilisées en hydraulique\rem{s}, mais ne sont à notre connaissance pas, ou très rarement, adapté\add{e}s aux modèles d'écoulement\rem{s} en couche\rem{s} mince\rem{s} sur des topographies quelconques, impliquant notamment de fortes pentes.


\subsection{Utilisation des modélisations pour l'alerte en temps réel}

Les modèles d'écoulement\rem{s} en couche mince\rem{s} permettent de réaliser des simulations avec des temps de calcul\rem{s} plus courts que des modèles 3D. Toutefois, ces temps de calcul \rem{sont couvent}\add{restent} trop importants pour envisager des modélisations en temps réel\rem{s}. Une possibilité pour utiliser les résultats des simulations pour de l'alerte est donc de construire une base de données de simulations pour différentes zones sources, volumes et paramètres rhéologiques. En fonction des informations disponibles pour préciser l'aléa (par exemple, le volume et la localisation des effondrements à partir d'enregistrements sismique\add{s} \citep[e.g.][]{durand_link_2018}, et/ou les prévisions de précipitations), les experts et acteurs pourraient alors avoir directement accès à la carte de propagation correspondante. 

 \citet{peruzzetto_operational_2020} proposent une démarche de ce type pour estimer les distances de parcours, en déterminant des lois puissances spécifiques à trois sites d'étude, donnant la distance de parcours en fonction du volume des glissements. Des méthodes plus poussées existent pour extrapoler les résultats de simulations, pour des paramètres d'entrée\rem{s} non testés dans la base de données de simulations. \citet{rohmer_dynamic_2014} analyse ainsi avec des "méta-modèles" la dynamique d'un glissement de terrain en fonction de variations de \add{la} pression de pore\rem{s}. Plus récemment, \citet{navarro_surrogatebased_2018} utilisent une approche similaire pour calibrer des simulations de coulées de débris.
 
 \rem{Ces méthodes ne sont néanmoins pas utilisées, à notre connaissance, pour des applications opérationnelles de surveillance et d'alerte. C'est pourtant le cas pour les tsunamis \citep{titov_realtime_2005}, ce qui souligne l'intérêt de poursuivre les développements dans cette direction.}\add{Ces méthodes pourraient potentiellement être utilisées pour des applications opérationnelles de surveillance et d'alerte, comme c'est le cas actuellement pour les tsunamis \citep{titov_realtime_2005}. Toutefois des développements méthodologiques restent nécessaires du fait de la complexité et des processus contrôlant le déclenchement et la propagation des mouvements de terrain.} 
 
\section{Conclusion}

Les modèles d'écoulement en couche mince permettent de quantifier la propagation des écoulements gravitaires. A l'échelle d'un site où les zones sources sont identifiées, ils offrent le bon équilibre entre facilité d'utilisation (temps de calcul aceptable, peu de paramètres à calibrer) et précision des résultats (épaisseurs et vitesses de l'écoulement en fonction du temps). En particulier, le code SHALTOP utilise des rhéologies simples, semi-empiriques, mais décrit finement les interactions géométriques entre l'écoulement et la topographie\rem{, ce qui est important} pour modéliser les interactions avec des obstacles. SHALTOP permet ainsi de reproduire les caractéristiques principales d'écoulements gravitaires secs (e.g. avalanche\add{s} de blocs) et hydro-gravitaires non visqueux (e.g. coulées de débris granulaires). \rem{Ces résultats ont été obtenus en récupérant et en exploitant  finement des}\add{Cela n'est toutefois possible qu'après la récupération et l'analyse de} données de terrain pour calibrer le modèle et définir des scénarios de simulation réalistes.

Il existe de nombreuses pistes de recherches pour améliorer et développer l'utilisation opérationnelle des modèles d'écoulement en couche mince\rem{. Par exemple, un travail important est}\add{, comme} le développement des modèles pour intégrer de manière robuste et fiable des processus \rem{importants }comme l'érosion. D'autres axes de recherches concernent les développements méthodologiques pour la cartographie de l'aléa et des risques, et l'utilisation des modèles pour des systèmes d'alerte et de surveillance en temps réel.

\section{Remerciements}

Les auteurs remercient le Ministère de la Transition Ecologique et Solidaire, le BRGM, L'Université de Paris (ANR-18-IDEX-0001), l'Institut de Physique du Globe de Paris (IPGP), la DEAL Martinique et l'ERC (ERC-CG-2013-PE10-617472 SLIDEQUAKES) pour le financement des travaux présentés ici. Les équipes du BRGM de Martinique, du BRGM de Guadeloupe, et de l'Observatoire Volcanologique et Sismologique de Martinique de l'IPGP ont également contribué à la collecte et à l'interprétation des données de terrain dans la Rivière du Prêcheur. 

\bibliographystyle{apalike}
\bibliography{references_bibtex}

\end{document}